\documentclass{elsart}

\usepackage[dvips]{graphicx}
\usepackage[dvips]{color}
\usepackage{refmerge}
\usepackage{latexsym}
\usepackage{amsmath}
\usepackage{amssymb}

\begin{document}

\begin{center}
\hfill  Preprint BINP 2003-25
\vspace {1cm}

{\Large \bf Study of the radiation hardness of CsI(Tl)
scintillation crystals }
\\[1cm]
\renewcommand{\thefootnote}{\fnsymbol{footnote}}
{ D.M.Beylin, A.I.Korchagin, A.S.Kuzmin, \\[1mm]
L.M.Kurdadze, S.B.Oreshkin, S.E.Petrov, B.A.Shwartz
\footnote{contact author, e-mail: shwartz@inp.nsk.su}}
\\[3mm]
{\it Budker Institute of Nuclear Physics\\
630090 Novosibirsk, Russia}
\\[3cm]
{\bf Abstract}
\end{center}

This paper is devoted to the study of a degradation of CsI(Tl)
crystals scintillation characteristics under  irradiation
with $\gamma$-quanta at the uniformly distributed
absorbed dose up to 3700 rad.
The sample set consisted of 25 crystals of 30 cm long
having a truncated pyramid shape and 30 rectangular crystals
of the same length. 
A large difference in the light output deterioration  
caused by the radiation was observed for the samples of the same shape.
A substantial dependence
of the average light output loss from the sample shape is seen as well.
On the other hand, the crystals from the same ingot behave very similarly
under irradiation.


\pagebreak

\section{Introduction}

Scintillating crystals of cesium iodide are widely used 
in high energy physics for photon detection. 
Calorimeters made of CsI crystals achieve the best energy resolution 
for photons and electrons \cite{ktev,ecl}.

One of the important characteristics of scintillating materials is their 
radiation hardness, i.e. their ability to retain scintillation efficiency 
and uniformity over the crystal volume after exposure to ionizing 
radiation. The radiation hardness of CsI crystals has been previously
studied in \cite{koba,renker,tit-rad,rad99}. 
Each of these studies was performed with a set of few crystals 
of different sizes and shapes, produced by different manufacturers. 
Despite the large spread of the results, it could be deduced that 
the CsI crystals keep the scintillating properties at the
acceptable level  
after the absorption of a radiation dose up to a few hundred rads. 
Also, 
it was clear, that the degradation of scintillating properties of 
crystals depends significantly on growing technology, raw material 
quality and size of the crystals. 

The interest to the radiation hardness 
of this material increased with the development and construction of
B-, $\phi$-, $c-\tau$ factories, e.g 
 storage rings with high luminosity which implies high 
electron and positron currents, 
causing high  radiation dose absorbed by the detector components.
Recently, this subject became topical again due to the development 
of new B-factory projects with super high luminosity \cite{sbelle,sbabar}.

This study was motivated by an active participation of 
BINP in the development and construction of the electromagnetic 
calorimeter of 
the BELLE detector \cite{belle}, in operation now at the 
KEKB collider  with luminosity of 
$10^{34}$~cm$^{-2}$s$^{-1}$. 
The electromagnetic calorimeter of this detector contains 
8736 crystals of CsI (Tl). Crystals are truncated pyramids of 
300 mm length  and transverse dimensions of 50-60mm. About 2/3 
of the calorimeter elements were produced by BINP in collaboration 
with the Institute for Single Crystals of National Academy of 
Sciences of Ukraine (Kharkov). To achieve and retain high energy 
and spatial resolution of the calorimeter it is important that 
the light output and uniformity of the crystals do not change significantly 
over a few years of the experiments at the collider. 
Therefore, crystals had 
to pass rather high requirements on radiation hardness.

During initial stages of the calorimeter development, the studies of 
the radiation hardness of crystals were performed with a few samples
\cite{oresh,tit-rad}. 
These studies demonstrated that crystals satisfy the requirements.

This paper is devoted to study the radiation hardness of 
CsI(Tl) crystals using a large number of full size samples. 
All 55 studied crystals were grown using the same technology 
by Joint Enterprise "Amcrys-H" of Scientific and Technological Concern
 "Institute for Single Crystals", Kharkov, Ukraine.
 Raw material for all crystals, high purity CsI, was supplied 
by Novosibirsk Plant of Rare Metals. For irradiation of the samples we 
used wide bremsstrahlung $\gamma$-quanta beam, 
produced by the 1.4 MeV electron accelerator, ELV-6, at BINP SB RAS. 
 The absorbed dose was distributed nearly uniformly over the crystal 
 volume.
Another feature of this work 
is that the absorbed radiation dose were measured  
with the detector based on CsI(Tl), thus avoiding error-prone 
recalculation of this value depending on the radiation energy 
spectrum. Preliminary results were reported at the INSTR-99 
conference \cite{instr}. 

\section{Studied samples and measurements of their 
scintillation properties}\label{samples}

From 55 studied samples, 25 were selected from crystals produced for 
the electromagnetic calorimeter of the BELLE detector. 
These crystals failed stringent 
geometry specifications or had small mechanical defects but
their scintillation properties met requirements. 
The samples, 
further referred as type "B", have a shape of truncated pyramid of 
300 mm height and average transverse sizes of about 60x60 mm$^2$. 
Detailed description of crystals used in the Belle detector
is given in \cite{belle_tdr}. 
The studied set comprised the crystals of 16 types which slightly 
vary in shape and 
sizes. Area of the large face of the truncated pyramid varies 
from 35 to 47 cm$^2$ while the angle between the pyramid sides and 
the axis ranged from 0.8$^{\circ}$ to 1.25$^{\circ}$, 
depending on the crystal type.

Other 30 studied samples were shaped as rectangular parallelepiped 
with sizes of $ 2\times2\times30$ cm$^3$. These crystals were specially 
manufactured for the systematic control of the radiation hardness
 of the calorimeter elements. 
These crystals are referred to as crystals of type "P".

Prior to measurements all crystals were polished, wrapped in 200 $\mu$m 
porous teflon and covered with 
20 $\mu$m thick aluminized mylar film.

The measurements of the light output of the crystals of type "P"
can be done with 
the photo detector attached to any of two edges of the sample. 
The light output depends on whether the other 
end is covered with reflector or not. 
Marking one end of the crystal as {\bf a}, 
and the another one as {\bf b}, we classify 4 possible options 
of the measurements of each 
crystal in the Table \ref{ab1}.
\begin{table}
  \centering
\caption{Four options of the light output measurements for
crystals of type "P" }
\begin{tabular}{|c|c|c|}
\hline
Method & Light is measured on  & 
Conditions on the Other End \\ \hline
 1 &End {\bf a} & End {\bf b} is covered with 2 layers of Teflon \\ 
 2 &End {\bf a} & End {\bf b} is not covered \\
 3 &End {\bf b} & End {\bf a} is covered with 2 layers of Teflon \\
 4 &End {\bf b} & End {\bf a} is not covered \\ \hline
\end{tabular}
\label{ab1}
\end{table}
In this work most measurements were performed using 
method 1. This method is implied throughout the text unless otherwise stated. 
For all measurements the sides of the crystal were covered by teflon and 
mylar in the same manner as for the samples of type "B". 

\begin{figure}[hbtp]
  \centering
  \includegraphics[width=10cm]{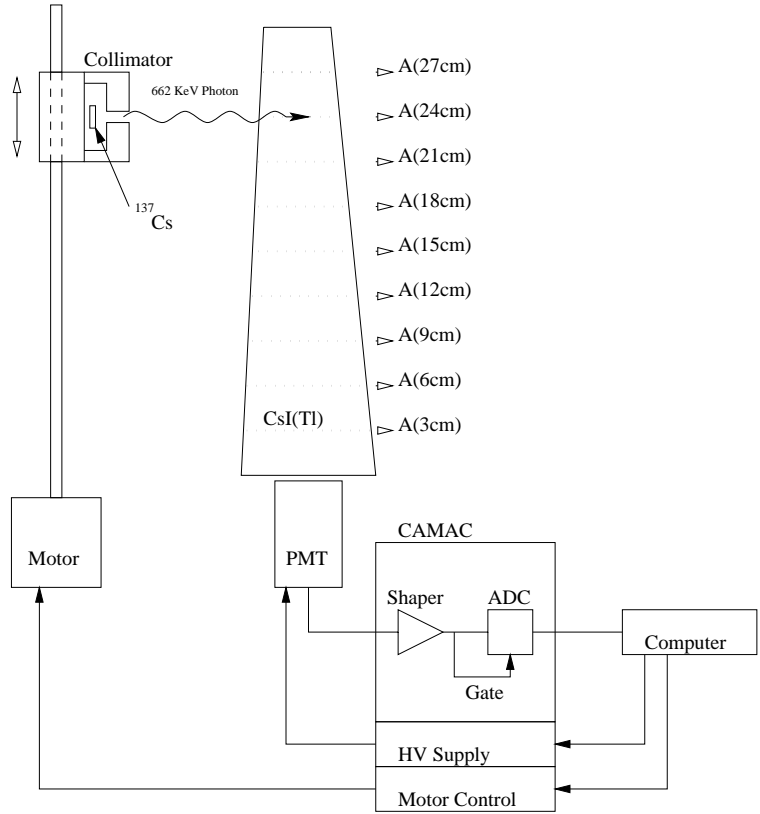}
  \caption{The light output measurement setup.}\label{stend}
\end{figure}

To measure the scintillation properties of the crystals we used the setup 
shown on Fig.~\ref{stend}. Studied crystal is placed vertically on the input 
window of the two inch PM tube (Hamamatsu R1847-07) with the photocathode 
without optical contact. 
Collimated $^{137}$Cs radioactive source emits 662 keV 
photons irradiating about 1 cm wide band of the studied crystal.
Signal from the PMT was shaped with shaping amplifier, digitized with 
ADC and transmitted to the computer.

Collimator can be moved along the axis of the crystal, 
irradiating the crystal with $\gamma$-quanta at 9 positions 
every 30 mm. For each source position the pulse height
spectrum was measured and the total absorbtion position was
determined. Then the light output was defined as a ratio of this
value($A_i$) and the corresponding value for the reference
crystal ($A_0$):
\begin{equation}
   L_i = {     {A_i}\over{A_0}   } \times100\%. 
\end{equation}
Average light output was defined as:
\begin{equation}
   \overline{L} = {( {\sum\limits_{i=1}^{9} {L_i} )} / {9}       }.
\end{equation}
Non-uniformity of the crystal was defined as:
\begin{equation}
    G = {{L_{max} - L_{min}} \over{L}     } 
\label{nonuni}    
\end{equation}

The reference was a CsI(Tl) standard detector  of  25mm height
and  25mm diameter packed in the aluminum container.
Troughout this paper the light output is referred as a ratio to the
reference one. 

\begin{figure}[hbtp]
  \centering
  \includegraphics[width=0.45\textwidth]{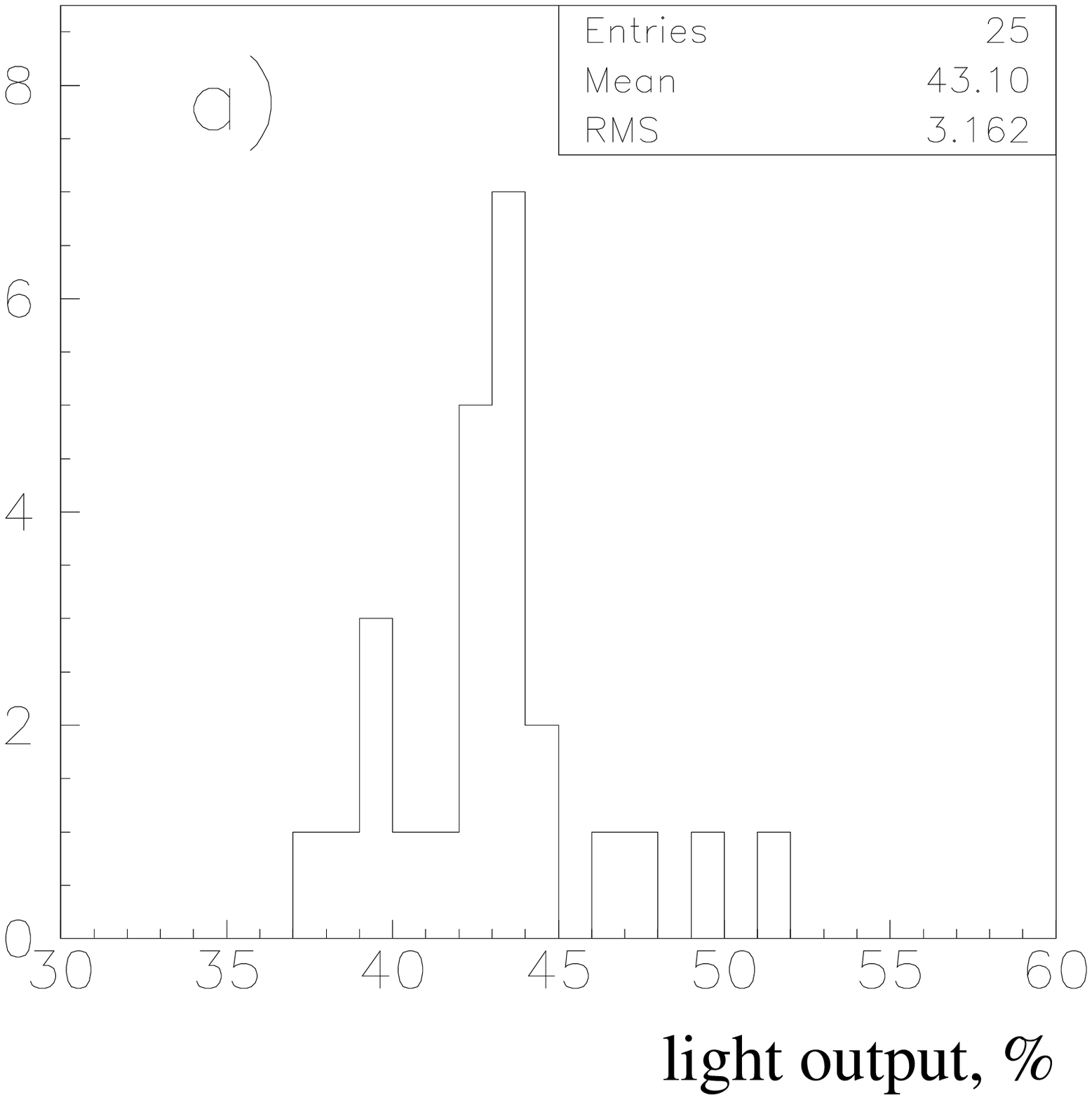}
  \hfill
  \includegraphics[width=0.45\textwidth]{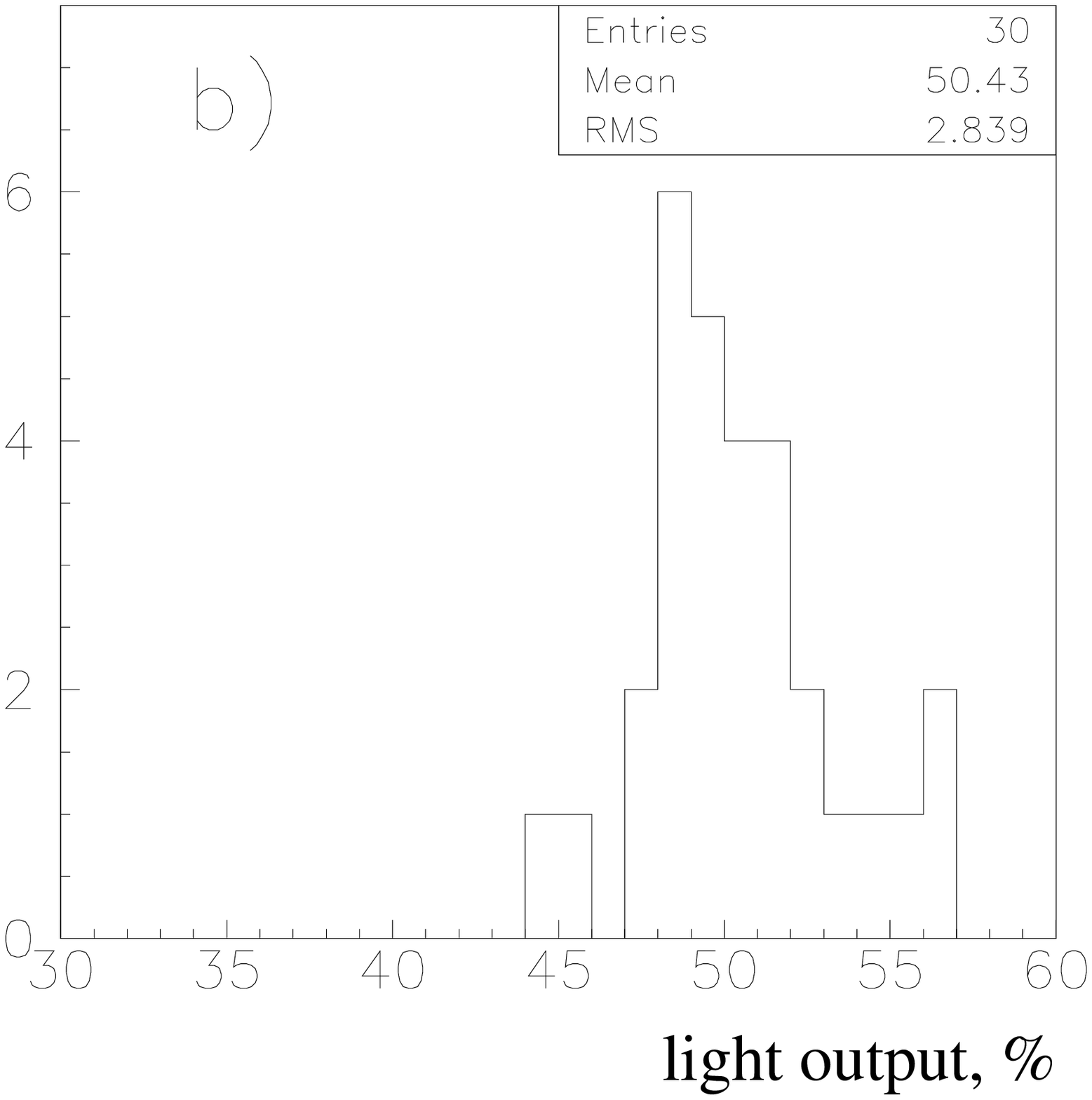}
  \caption{The light output of the crystals type "B" ($a$) and "P" ($b$)
  before the irradiation.}
  \label{piram_l}
\end{figure}

Fig.~\ref{piram_l},$a$ and $b$
represent the distribution of samples of type "B" and "P" 
over the value of the average light output, measured before 
the irradiation. It can be seen from the histograms that the 
initial spread of the light output  is rather small.

\section{Irradiation of the Crystals}\label{irrad}

Accelerator ELV-6 \cite{elv} generates continuous 1.4 MeV energy 
electron beam  with up to 100 mA current. The power in the beam 
reaches 100 kW. The beam with such parameters can produce 
radiation dose intensity of up to 1 Mrad/hour. 
This work studied the radiation hardness at moderate 
doses, and the irradiations were performed at a current that
did not exceed 100mA. 

\begin{figure}[hbtp]
  \begin{center}
    \includegraphics[totalheight=7cm]{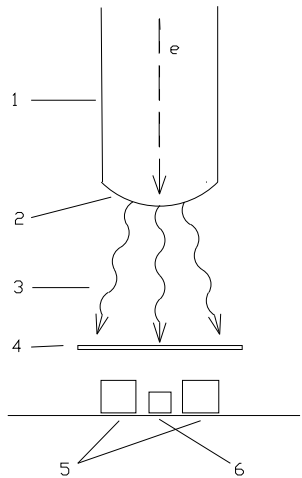}
  \end{center}
  \caption{The setup for irradiation of crystals with the 
   ELV-6  accelerator.
   1 -- accelerator of electrons ELV-6, 
   2 -- tantalum converter, 3 -- bremsstrahlung photons, 
   4 -- 2mm lead plate, 5 -- irradiated crystals of CsI(Tl), 
   6 -- detector of radiation dose.}
  \label{elv}
 \end{figure}

The scheme of the irradiation setup is shown in Fig.~\ref{elv}. 
The electron beam of 1.4 MeV energy  hits the converter (2), composed 
of 0.5mm Ta, 2mm water and 2mm stainless steel. 
Bremsstrahlung $\gamma$-quanta produced in the converter 
irradiate crystals (5) located under the beam 
with its axis perpendicular to the beam direction. Dose detector (6) 
is placed between the crystals. 2mm plate of  lead is used to suppress the 
low energy part of the photon spectrum.

The resulting photon beam has wide spectra from 0.2 to 1.4 MeV with maximum 
around 0.6 MeV.

The energy deposited by ionizing radiation in a unit mass of the 
material is absorbed dose (D). It is measured in units of 
1~Gy = 1 J/kg. Another 
unit of absorbed dose is 1~rad = 0.01 Gy (1~rad = 10$^{-2}$ 
J/kg = 5.24$\times$10$^8$ MeV/g).

The  absorbed dose is determined by the radiation intensity and 
spectra, as well as by the material of the sample. 
Radiation flux observed by samples
consists of the primary radiation and the photons 
scattered by the surrounding materials.
That makes calculation of the 
absorbed dose difficult even for monochromatic gamma-source. 
Detectors 
of absorbed dose are usually based on low-Z materials; this leads to 
significant systematic errors when re-calculating to CsI. Therefore, 
for the purpose of this study, we made the dose detector (DD) on the 
basis of scintillation crystal CsI(Tl). Using the CsI as a detector 
material we avoid dose recalculation and improve accuracy 
of absorbed dose measurement.

\begin{figure}[hbtp]
  \begin{center}
    \includegraphics[width=10cm]{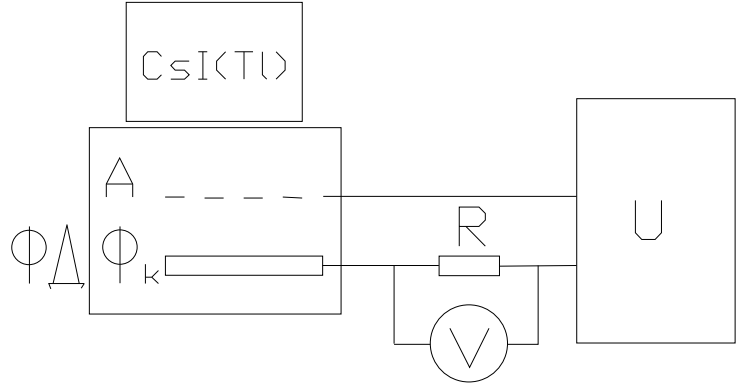}
  \end{center}
  \caption{ Detector of absorbed dose. CsI(Tl) -
    scintillation crystal, PD - Photodiode with massive photocathode, 
    A - Anode, C - photocathode, R - resistor, V - voltmeter, 
    U - DC power supply. }
  \label{izmer}
 \end{figure}

The layout of the DD is shown in Fig.~\ref{izmer}. 
Scintillation crystal CsI(Tl) 
with dimensions of 2$\times$2$\times$1 cm$^3$ is coupled to  
a vacuum photodiode (PD) with massive photocathode. 
Photocurrent is proportional to the dose rate
in the wide range of the radiation intensity. It is measured by 
the voltage drop (V) on the load resistor (R=50KOhm)

Detailed description of the DD design and calibration procedure is given 
in Appendix A. 
It was measured that the intensity of 
absorbed dose of 1~rad/sec results in a photocurrent 
of about 0.3$\mu A$. 
Then the   dose D(rad) absorbed
by the crystal during time $t_0$(s) at the current $I (\mu A)$ 
can be calculated as:
\[ D = \frac{1}{I_R} \int_{0}^{t_0}I(t)dt ,\]
 Where $I_R = 0.3\mu A\cdot sec/rad$.

\begin{figure}[hbtp]
  \centering
   \includegraphics[width=10cm]{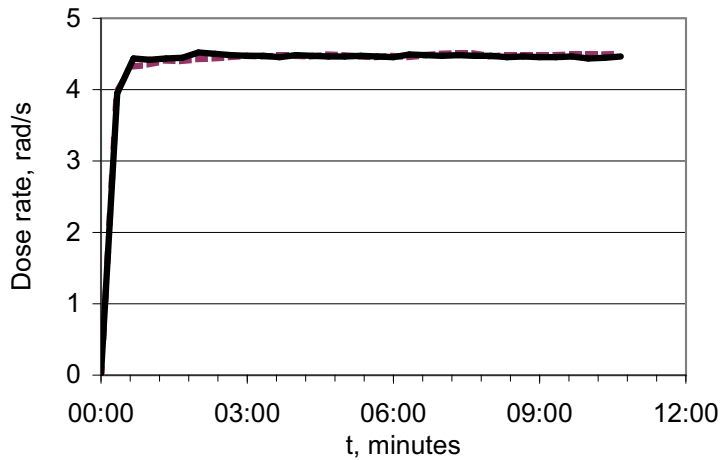}
  \caption{Typical dose rate in dependence of the time of exposition.
  Solid line corresponds to the irradiation of 
crystals from one side, dot line - from the other one (see text).}
\label{dose_t}
\end{figure}

During the irradiation, the current value was recorded every 20 sec. 
Typical time dependence of the dose rate calculated from the
DD current  is shown in the Fig.~\ref{dose_t}.

The relative accuracy of the absorbed dose measurement is about 10\% that 
is determined by the accuracy of DD calibration and taking into account 
slow scintillation components of CsI(Tl).

Simple estimations show that the multiple scattering of the 
electrons in the converter result in a good uniformity of the 
$\gamma$-beam over crystal location area.
To check this,  the transverse profile 
of the dose intensity at the fixed accelerator current was measured. 
Nonuniformity 
was found to be less than 5\% within the range from -15cm
to 15~cm from the electron beam axis.

The $\gamma$-beam used for irradiation has wide energy spectrum 
from 0.2 to 1.4~MeV with maximum around 0.6MeV. At this energy, the 
average photon interaction length
in CsI is about 3~cm. Since
the transverse size of the crystal is about 6cm 
the dose absorbed near the upper side 
of the crystal is a few times higher than that at the bottom one.

To determine the magnitude of this effect, we performed the following 
measurement. With certain accelerator current we measured the intensity 
of the dose absorbed by the DD (J1). Then the DD was shielded with a 6cm 
thick block of CsI and the measurement was repeated at the same accelerator 
current (yielding J2). The ratio k=J1/J2 was found to be equal to 3.2 which, 
assuming the exponential dependence of the absorbed dose on the depth, 
corresponds to the average $\gamma$-quanta attenuation length of 5.2cm.

\begin{figure}[hbtp]
  \centering
  \includegraphics{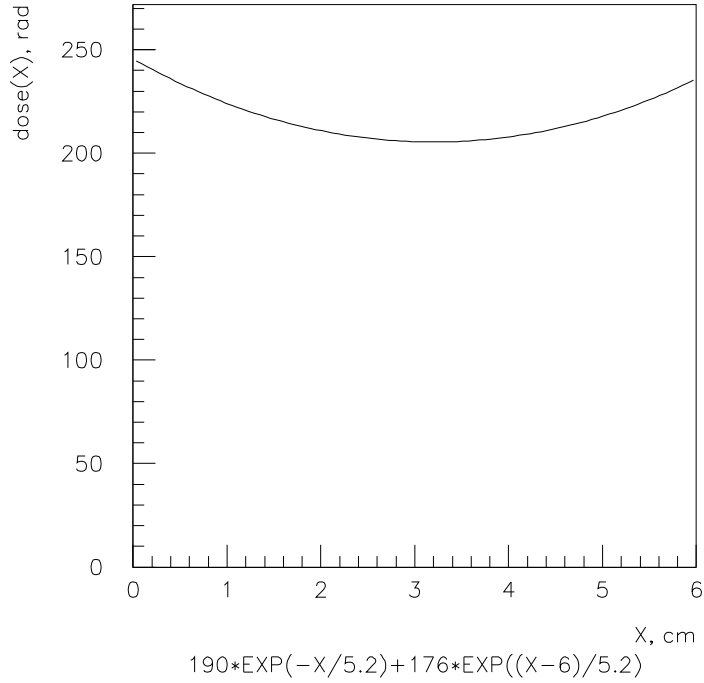}
  \caption{The distribution of the absorbed dose over the 
crystal depth after irradiation from both sides.}
\label{dose_s}
\end{figure}

To compensate this nonuniformity each sample was irradiated with 
equal doses from opposite sides. 
Fig.~\ref{dose_s} shows the dependence of the 
dose on the depth calculated for k=3.2. It can be seen from the picture 
that nonuniformity of the absorbed dose over the depth is about 15\%.

\section{ Changes of the crystal scintillation characteristics induced
by the radiation}\label{measure}

The light output of every crystal was measured several times before and 
after each irradiation. Irradiation sessions were carried out consecutively 
with 2-6 weeks intervals between them. 
No more than 3 samples of type "B" 
or 5 samples of type "P" were irradiated simultaneously in the same
session.

Significant afterglow of CsI(Tl) makes it difficult to measure the light 
output right after the irradiation. So, the first measurement 
was performed only one day after the irradiation when the afterglow  
became low enough 
not to broaden and shift the photo-peak significantly.

\begin{figure}[hbtp]
\vspace*{5mm}
  \centering
  \includegraphics[width=0.48\textwidth]{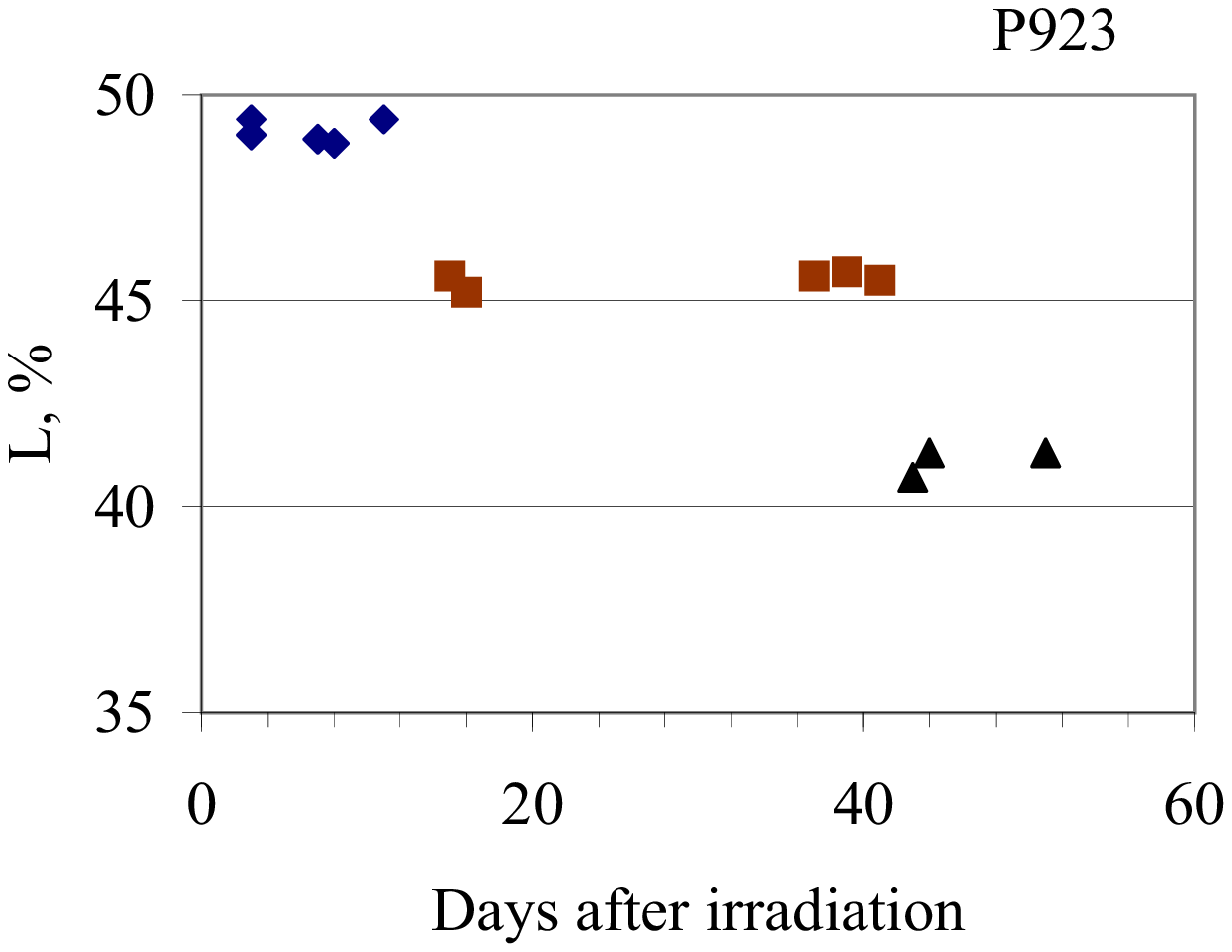}
  \hfill
  \includegraphics[width=0.48\textwidth]{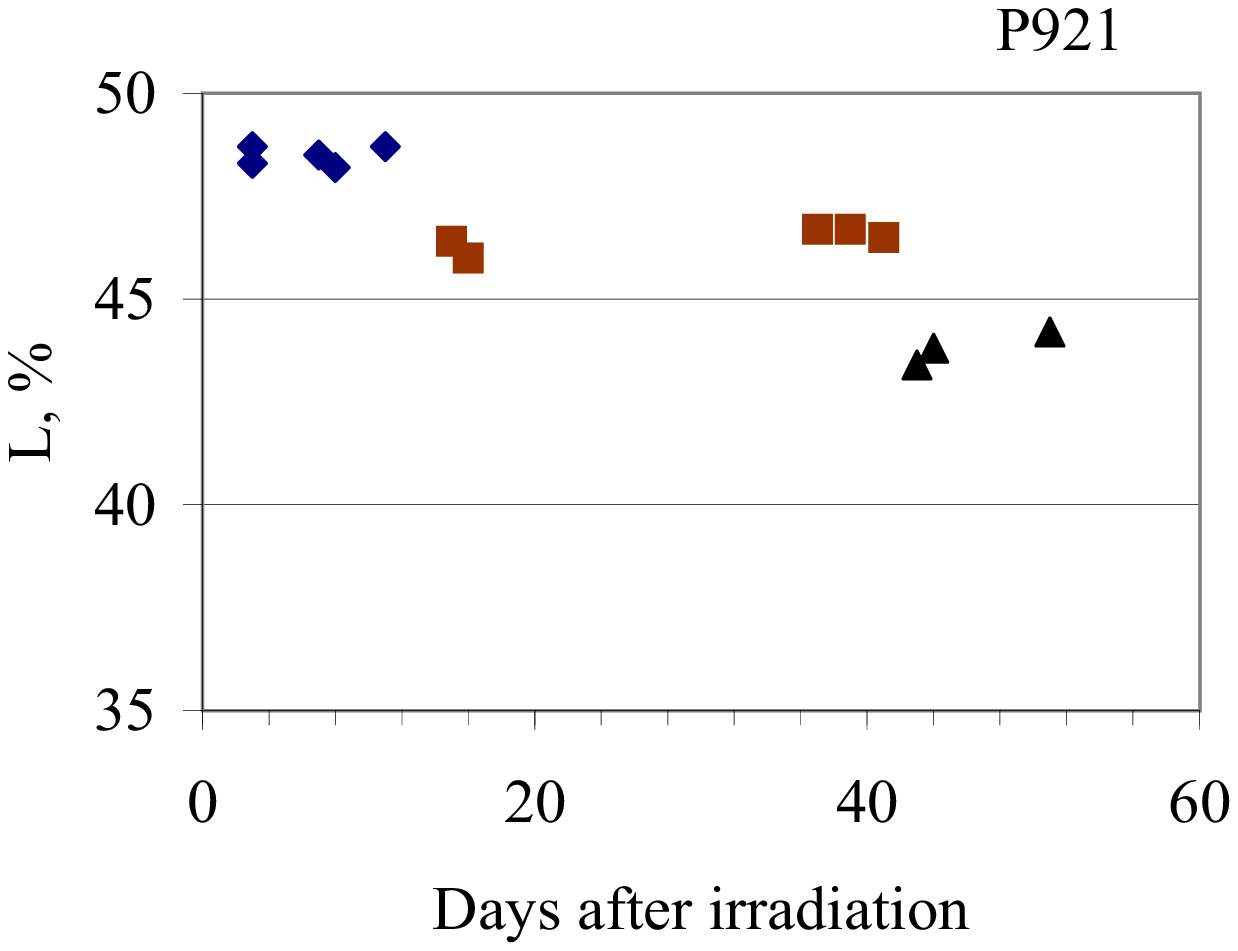}
  \caption{Light output of two crystals of type "P" as a function of time. 
  $\blacklozenge $ - light-output of the samples before the irradiation, 
  $\blacktriangle$ - after irradiation dose of 700 rad, 
  $\blacksquare$ - after 3200 rad. }
  \label{923}
\end{figure}

Fig.~\ref{923}
shows the average light output of two "P"-type crystals as a function 
of time. Stability of the measurements performed between the irradiations 
is better than 1\% which allows seeing sharp drops in the light output 
after irradiations with doses 700 and 3200 rad. Gradual increase in the 
light output after the second irradiation could be interpreted as a partial 
recovery of the light-output. This issue will be discussed in more details 
in the section \ref{part-rec}.

\subsection{ Measurements with the type "B" crystals (pyramids) }
\label{me_piram}

From the set of 25 samples shaped as truncated pyramids
nine crystals were irradiated 
several times up to the integrated dose of 3600-3700 rad, 
six crystals up to 1300-1400 rad, while other 10 samples were 
irradiated twice --- up to 740-780 rad and 3000-3200 rad.

\begin{figure}[htbp]
  \centering
  \includegraphics[width=0.48\textwidth]{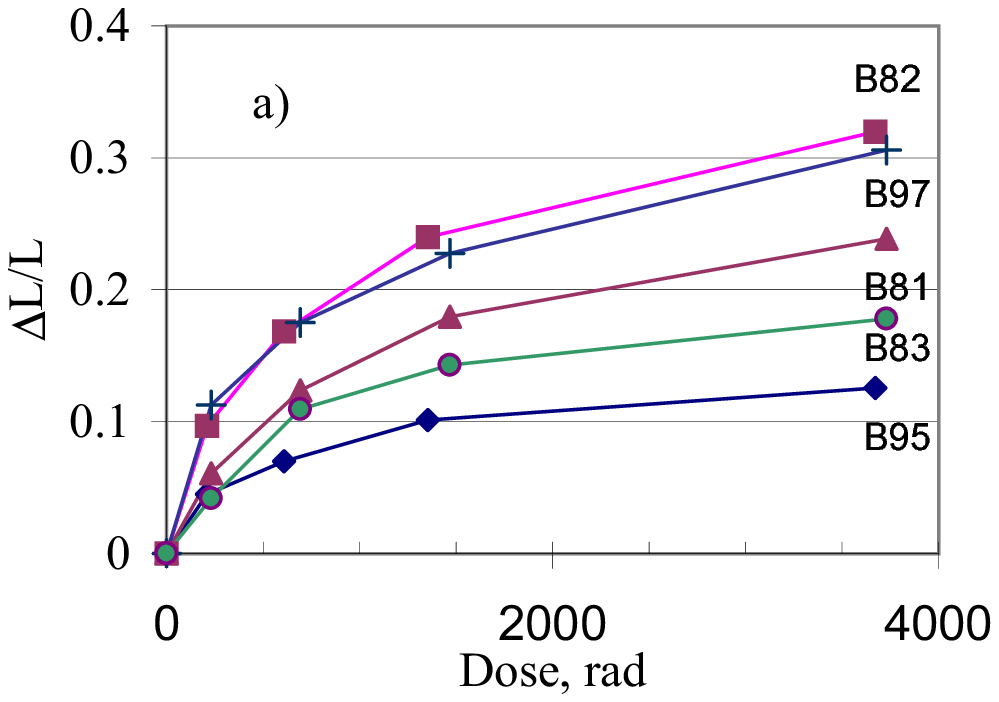}
  \hfill
  \includegraphics[width=0.48\textwidth]{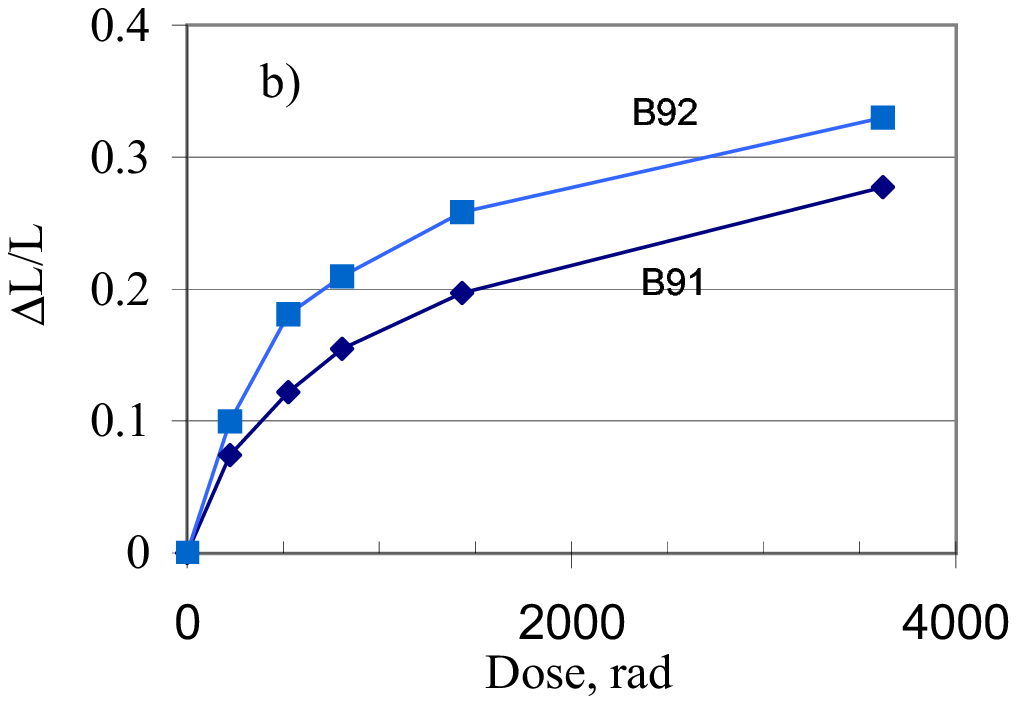}
  \caption{Light output degradation, $\Delta{L}/L$,
  in dependence of the absorbed dose for seven crystals
  of "B" type.}
\label{dl-l}
\end{figure}

Fig.~\ref{dl-l} shows the degradation of the light output,
$\Delta{L}/L$, as a function of the
absorbed dose for seven pyramid-shaped crystals (type "B"). 
As it's seen in 
the figure, all crystals are characterized by similar curves. The 
light output drops fast at the small irradiation doses but at higher 
doses it decreases slower. For example, for the crystal B91 the 
drop was 7\% after irradiation with 220rad, but only 27\% after increase 
of the dose up to 3.6 Krad.

\begin{figure}[htbp]
  \centering
  \includegraphics[width=\textwidth]{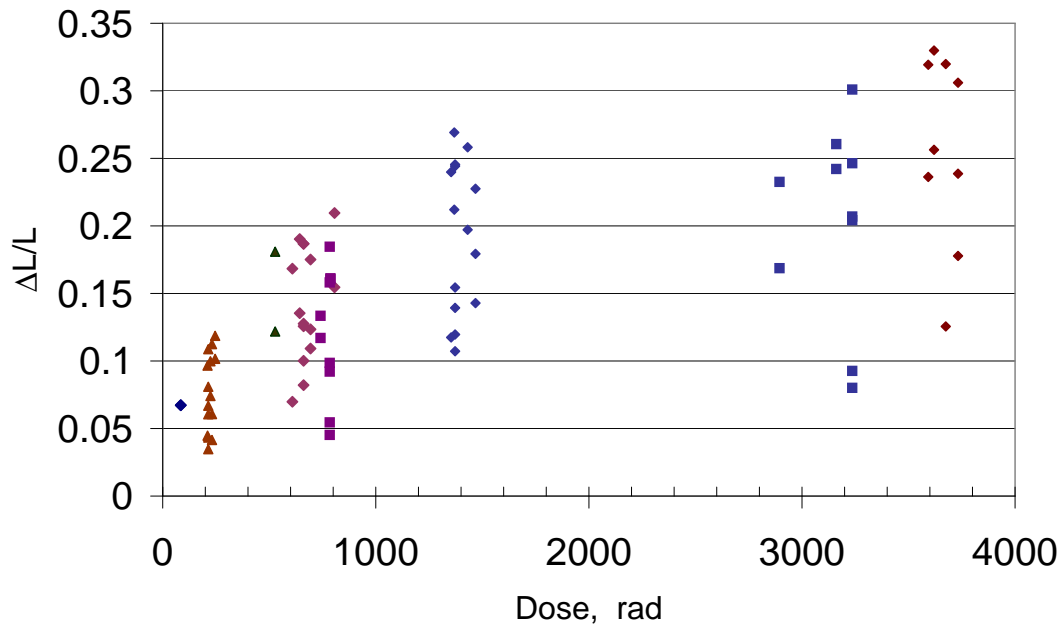}
  \caption{The light output decrease for all type "B" crystals after all 
irradiations. Samples of the "endcap" types are marked with -- $\square$.}
\label{bighist}
\end{figure}

Fig.~\ref{bighist}
shows the distribution of type "B" samples over the light-output 
loss, $\Delta{L}/L$, after all irradiations. It can be seen that the 
distribution is very wide. Thus, after irradiation with 700-800 rad, 
$\Delta{L}/L$ varies from 5\% to 21\%.

As it was mentioned earlier one source of raw material and one technology 
were used for the production of all studied crystals.
Therefore, the wide spread in $\Delta{L}/L$ is an evidence  
that the radiation hardness is, probably, strongly dependent on 
uncontrolled impurities in the raw material or slight variations in 
the crystal growth regime causing defects in the crystal lattice.

Production of crystals for the electromagnetic calorimeter of 
the Belle detector, 
from which the samples studied in this work were selected, took about 
four years. It took over three years to produce crystals for the barrel 
part of the calorimeter while all the crystals for the endcaps 
were produces in the last 8 months. 
One could suppose that over the long 
production period both growth technology and raw material could vary. 
In Fig.~\ref{bighist} 
the endcap type crystals produced near 
the end of the production are specially marked. 
As we can see from the picture the 
distributions over the $\Delta{L}/L$ for these crystals are very 
similar to those for the barrel ones.

\begin{figure}[htbp]
  \centering
  \includegraphics[width=10cm]{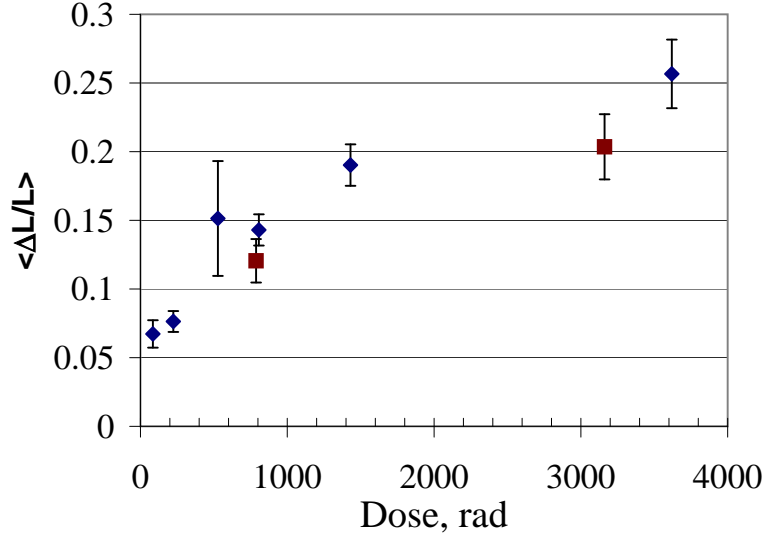}
  \caption{Average light output deterioration, $\Delta{L}/L$, 
  for the type "B" crystals as a function of the absorbed dose. 
  The "endcap" samples are marked with $\square $.}
\label{dl-ave}
\end{figure}

Fig.\ref{dl-ave} shows the average values of 
$\Delta{L}/L$ as a function of absorbed dose. 
Points corresponding to the "endcap" crystals are somewhat lower than 
the rest although the difference is insignificant when compared to 
the statistical errors.

Values of $\Delta{L}/L$ obtained in this study do not contradict 
to the results 
of the earlier measurements of the radiation hardness of a few samples 
for the electromagnetic calorimeter of BELLE produced by Crysmatec, 
Shanghai Institute of Ceramics (SIC) and Kharkov Institute for Single 
Crystals [5].

\begin{figure}[hbtp]
  \centering
  \includegraphics[width=0.49\textwidth]{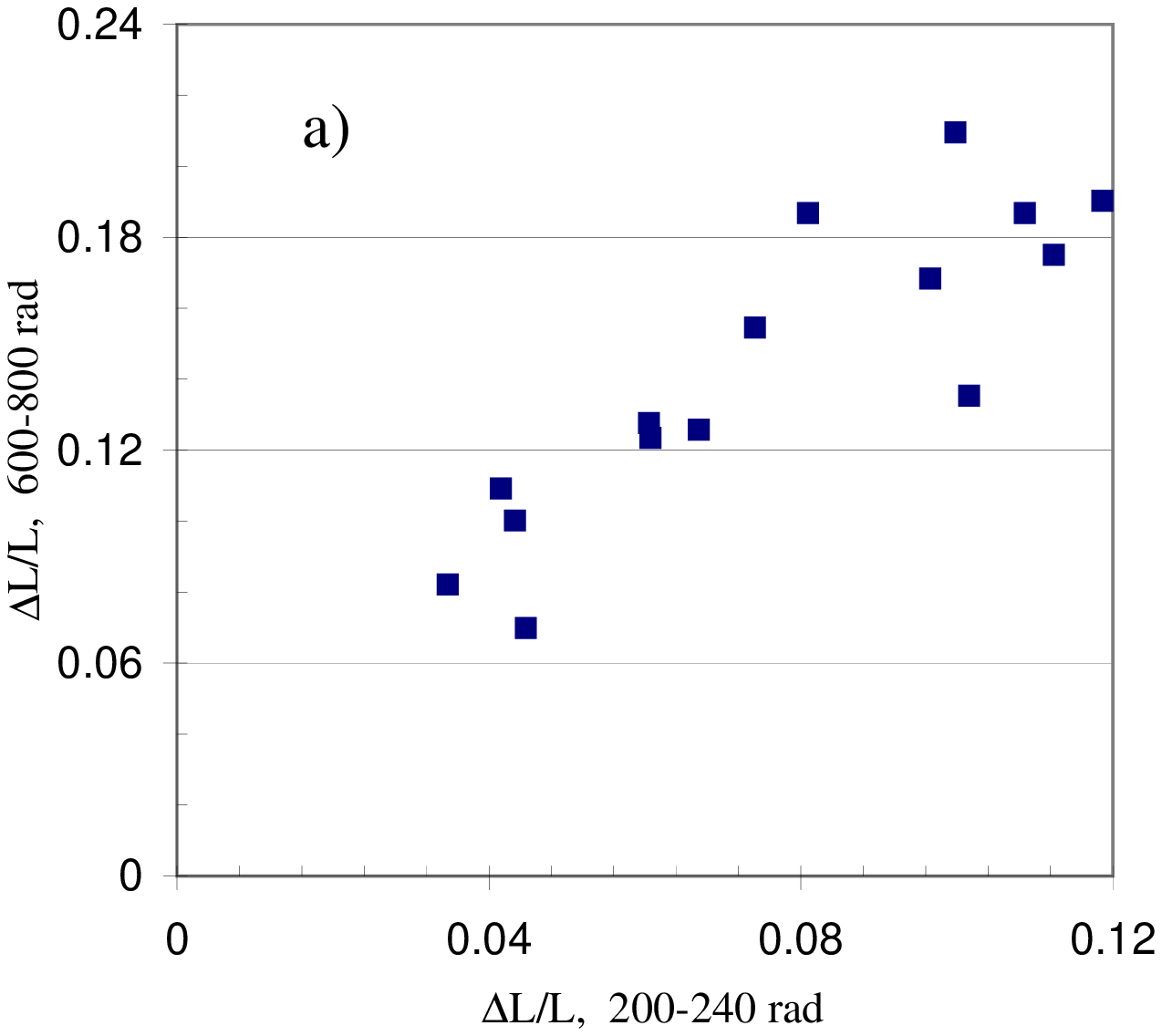}
  \hfill
  \includegraphics[width=0.49\textwidth]{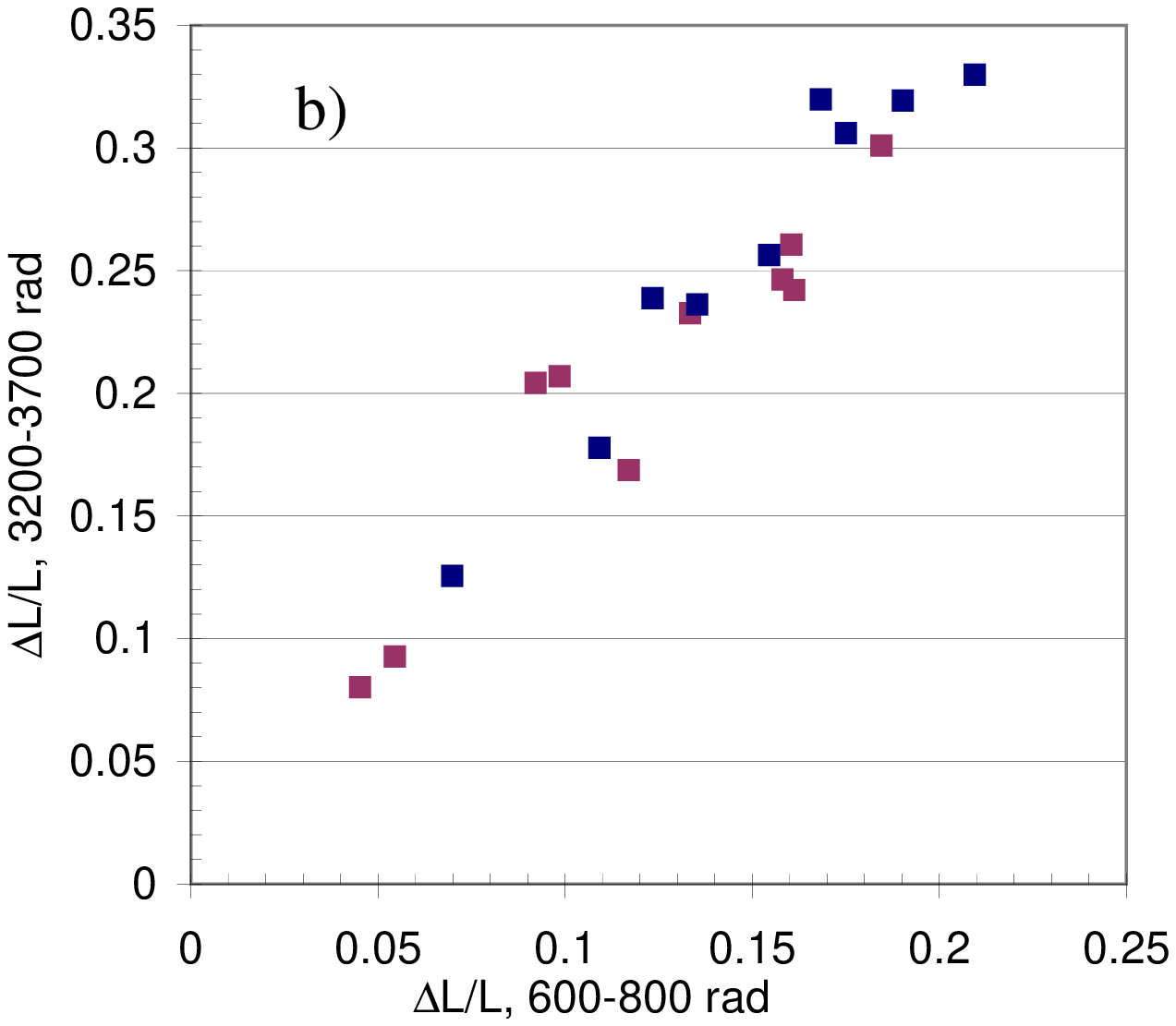}
  \caption{Correlation between the $\Delta{L}/L$ for 
  "B" crystals after the irradiation with integrated dose 
  reaches 200$\div$240 and 600$\div$800 rad ($a$), 
  600$\div$800 and 3200$\div$3700 rad ($b$). }
\label{cor1}
\end{figure}

We estimate the accuracy of the light output measurements 
in this work by about 1\%.
Therefore, the width of the distribution over $\Delta{L}/L$ at a given 
absorbed dose (see Fig.~\ref{bighist}) is not determined by the 
measurement errors.
This conclusion is supported by  a good correlation between the 
decreases of the light output of the same crystal after irradiation 
with different doses. Such correlation is presented in Fig.\ref{cor1}.

Strong positive correlation which can be seen in Fig.~\ref{cor1} 
supports the conclusion 
made earlier that the $\Delta{L}/L$ dependence on the absorbed 
dose has similar shape for all crystals.

The observed correlation makes possible the procedure of  
the radiation hardness control  for each crystal individually. 
By measuring 
the decrease of the light output after irradiation with a relatively 
small dose and further extrapolating to higher doses, one could come to 
an expectation about the ability of the crystal to withstand 
required radiation load.

Another conclusion can be made from the obtained results that if the 
crystals are to work under high radiation background when radiation 
hardness is important, it is necessary to study the radiation hardness 
of sufficiently large set of crystals. 

Since this work was initiated by the Belle detector construction  
it should be noted that all the studied samples satisfied the radiation 
hardness specifications required for the elements of the calorimeter.

It should be noted as well, that some of the studied samples 
demonstrated very 
high for the alcali halide crystals radiation hardness 
($\Delta{L}/L$=8\% at 3.2 Krad). It indicates that
the production of the radiation hard CsI crystals is possible
after improvements of the growing technology.

\subsection{ Measurements with the type "P" crystals (parallelepiped) }
\label{me_paral}

The set of "P" crystals contained 30 rectangular samples 
of $2\times2\times30$ cm$^3$ size. 25 of them were irradiated with
650 rad dose in one exposition and later with additional dose
of 2550 rad that resulted in the integrated dose of 3200 rad. 5 crystals
were irradiated only once with the dose of 3200 rad.

\begin{figure}
  \centering
  \includegraphics[width=10cm]{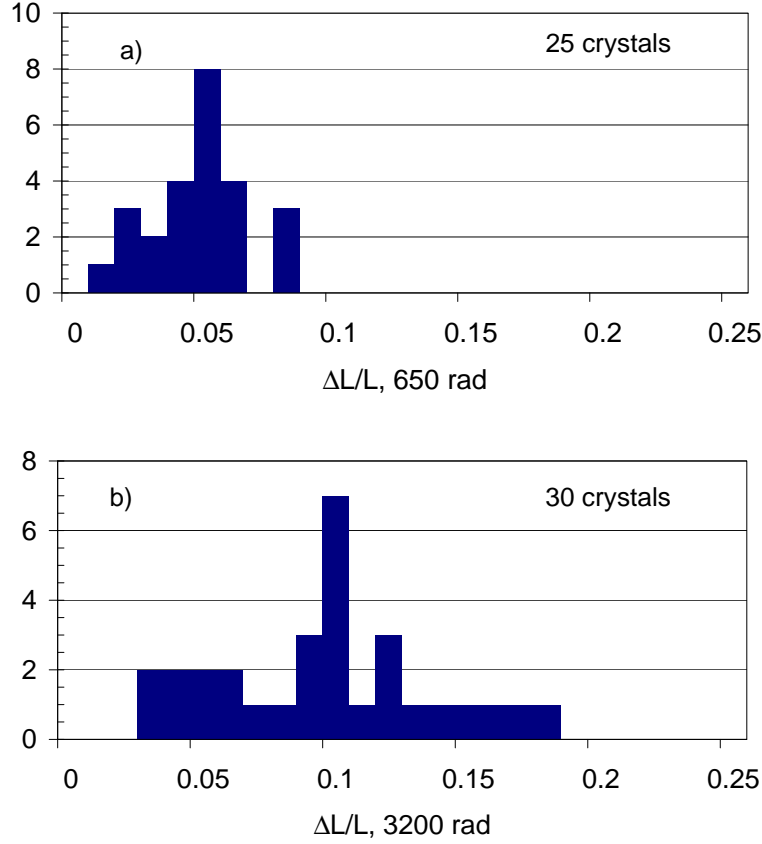}
  \caption{The distributions of "P" samples over light output 
  deterioration
    after the irradiation with 650 rad (a) 
   and 3200 rad (b).}
\label{para}
\end{figure}

The distributions over $\Delta{L}/L$ for "P" samples after the 
irradiation with 650 and 3200 rad are shown in Fig.~\ref{para}.

The distribution is wide like for the "B" crystals case. 
However, the average decrease is much less ---
$<\Delta{L}/L> =(4.2 \pm 1.8)\%$ at 650 rad and
$<\Delta{L}/L> =(9.0 \pm 3.9)\%$ at 3200 rad.
Comparison of these results with those shown in Fig.~\ref{dl-ave}
leads to a conclusion that the light output decrease for
parallelepiped shaped crystals is 2-3 times less that 
that for pyramid ones. 
This effect will be discussed in more details in the section \ref{dis}.

\begin{figure}
  \centering
  \includegraphics[width=10cm]{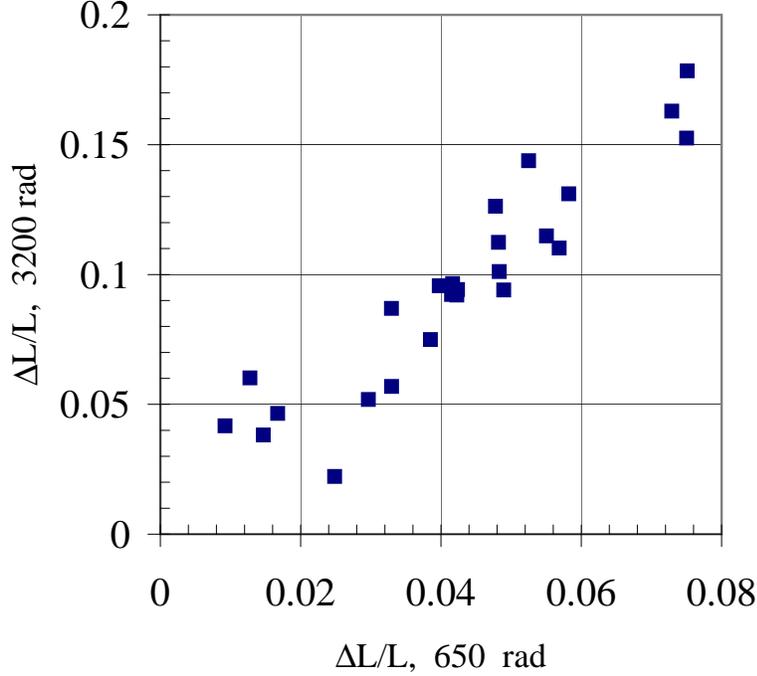}
  \caption{The correlation between the light output decrease 
  for 25 crystals "P" after the irradiation with 650 rad 
  and 3200 rad.}
\label{para1}
\end{figure}

The correlation between the light output decrease after the
irradiation with 650 rad and 3200 rad is shown in Fig.~\ref{para1}.
Similarly to the crystals "B", good correlation between 
$\Delta{L}/L$ values after irradiation with various doses 
is seen.

\subsection{ The light output change for type "P" crystals
at the various light collection conditions }
\label{me_vary}

\begin{figure}
  \centering
  \includegraphics[width=10cm]{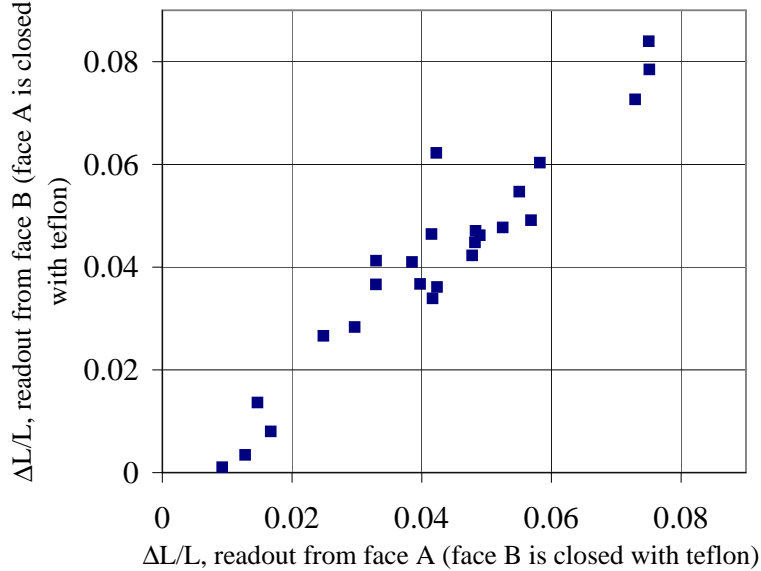}
  \caption{The correlation between $\Delta{L}/L$ values
measured from the different faces of P-type crystals
with options 1 and 3 (see Table~\ref{ab1} for explanation).
The absorbed dose is 650 rad.}
\label{ab}
\end{figure}

As it was mentioned in the section \ref{samples} there are
several options for light read out for "P" crystals
(see Table~\ref{ab1}). To estimate the precision of the $\Delta{L}/L$
measurement and the influence of the uncontrolled variations 
in the polishing of the crystal faces as well as wrapping with
teflon, we compared the light output after 650 rad dose irradiation
measured by option 1 (light is read from the edge {\bf a} while 
the edge {\bf b} is covered with two layers of teflon ) and option 3 
(light is read from the edge {\bf b} while the edge {\bf a} 
is covered with two layers of teflon).
The results are presented in Fig.~\ref{ab}. Clear correlation 
between the measurements performed by two options leads to the
conclusion that the uncontrolled variations in the procedure
of the detector preparation do not contribute substantially to
our measurements.

\begin{figure}
  \centering
  \includegraphics[width=10cm]{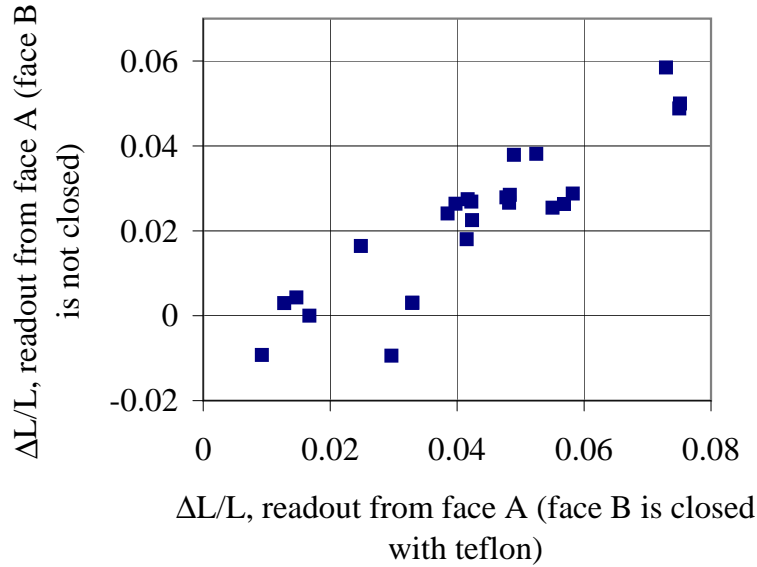}
  \caption{Correlation between the light output deterioration for
  crystals of "P" type at the measurement with open and 
  covered rare edge. The absorbed dose is 650 rad.}
\label{edge}
\end{figure}

Fig.~\ref{edge} shows a comparison of $\Delta{L}/L$ values
for 25 samples of "P" type measured after 650 rad dose irradiation
by option 1 (light is read from the edge {\bf a} while 
the edge {\bf b} is covered with two teflon layers) and
option 2 (light is read from the edge {\bf a} while 
the edge {\bf b} is open). Good correspondence between
the results obtained in both options is seen. The light output
deterioration with the open rare edge somewhat less than that 
with covered one.

\section{ Radiation hardness of the crystals produced from the
same ingot} \label{bul}

Among the samples of "B" type two pairs of crystals were produced from
the same ingot: samples B62 and B67 were cut from one ingot,
while both samples B65 and B64 were made from the another one.

\begin{figure}[htbp]
  \centering
  \includegraphics[width=0.48\textwidth]{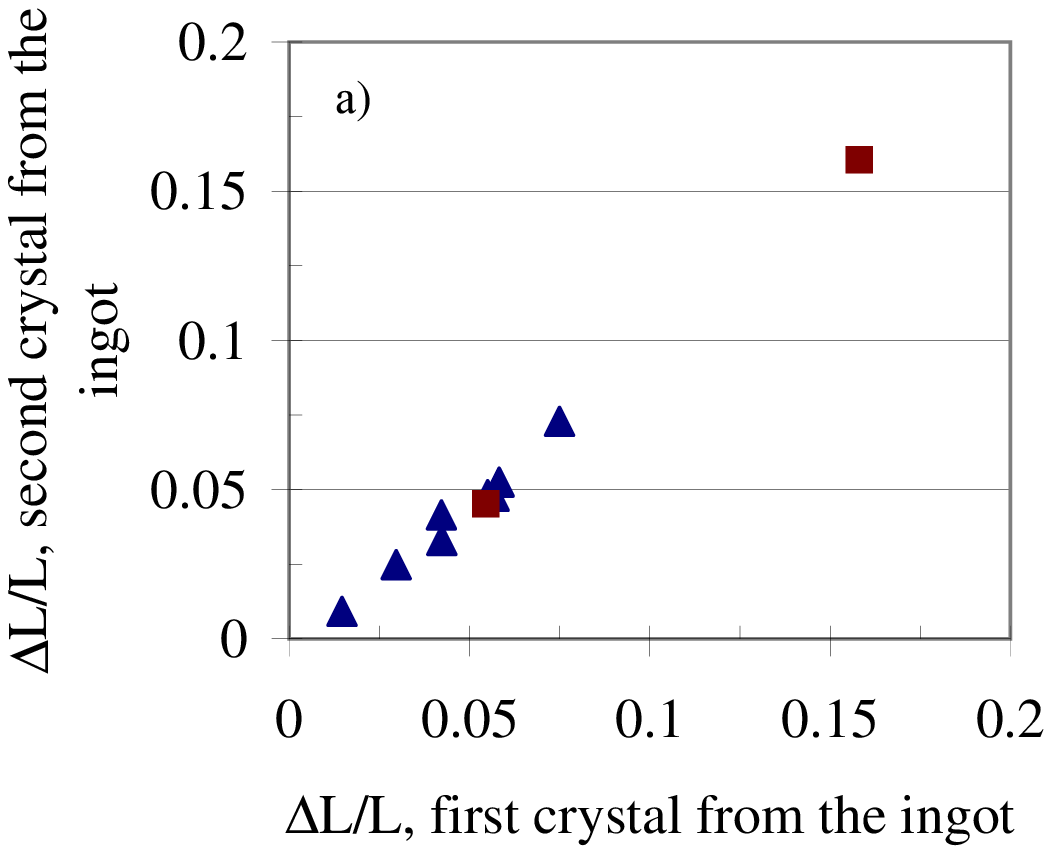}
  \hfill
  \includegraphics[width=0.48\textwidth]{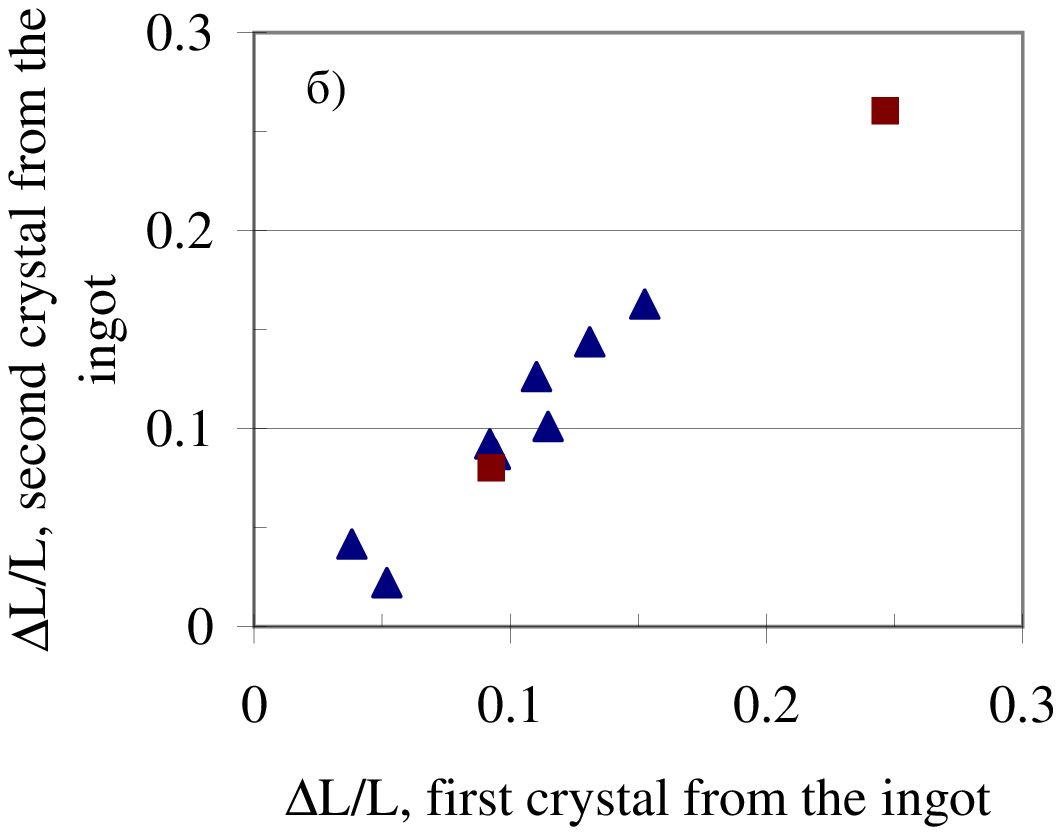}
  \caption{Correlation between the light output decrease for 10
  pairs of samples of both types, produced from the same ingot.
  The crystals were irradiated with dose 650-780 rad (a) and
  3200 rad (b).  $\blacktriangle$ -- "P" crystals, $\blacksquare$ -- 
  "B" samples.}
\label{para2}
\end{figure}

The set of "P" crystals contained 8 pairs of such type. In the plot of
Fig.~\ref{para2} the X-axis corresponds to $\Delta{L}/L$ value for 
the first crystal from the pair while this value for the second
crystal of the same pair is shown along the Y-axis.
Presented results were taken for all 10 pairs of crystals after
irradiation with the dose of 650-780 rad (a) and dose of 3200 rad (b).
As it's seen in the figure the radiation hardness of the crystals
produced from the same ingot is the same within the measurement
accuracy both for "B" type samples and for "P" ones.

Thus, one has to conclude that the light output deterioration of
the irradiated crystals is determined mostly by the properties
of grown ingot while the variation of the material properties over 
ingot volume as well as the uncontrollable changes of the 
procedure of the crystal processing make much smaller
contribution.

The observed correlation provides a possibility for mass control
of the radiation hardness of the detectors. It's enough for
that to irradiate one sample from each ingot with the desirable
dose.

\section{ Light output self-recovery after irradiation } 
\label{part-rec}

Some scintillation crystal manifest the natural recovery  
of the light output damaged due to irradiation. This process 
was observed for BGO  \cite{bgo-rec}, BaF$_2$ \cite{baf2-rec}, 
PbWO$_4$ \cite{pwo-rec} and for some other materials.
Existence of such a recovery for cesium iodide crystals 
was noted in the works \cite{tit-rad,csi-rec}.

To study this effect the special measurements
with 5 samples of "P" type were performed. The light output and
nonuniformity were measured daily during several days before 
irradiation and after one of 3200 rad. However, it should be remarked
that the recovery observation for CsI(Tl) crystals is complicated 
due to high afterglow that makes the measurements rather inaccurate in the
first day after the irradiation.

\begin{figure}[hbtp]
  \centering
  \includegraphics[width=0.7\textwidth]{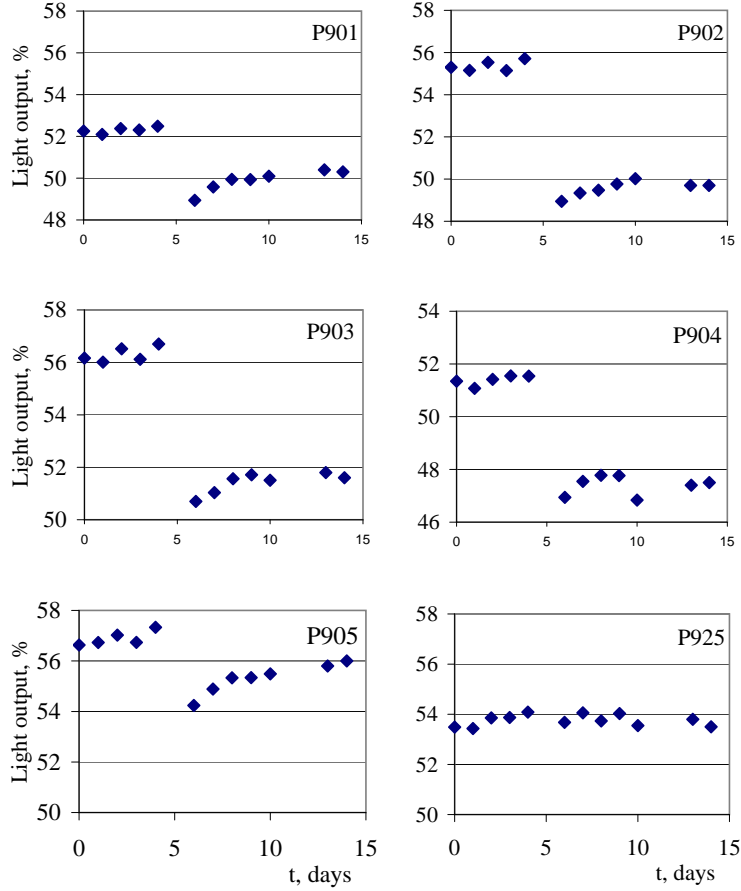}
  \caption{Time dependence of the crystals light output.
  Sharp change is connected with the exposition of the crystal with
  the dose of 3200rad. The results for P925 nonirradiated sample
  are referred for comparison}
\label{recov}
\end{figure}

Fig.~\ref{recov} presents light output time dependence before and
after irradiation for five crystals (P901 - P905). The results
for one crystal (P925) which was not irradiated is presented
as well for comparison.
It's seen that the light output is being recovered with a time
constant of a few days up to 20\% of the initial radiation damage.  

However, the reliable conclusion on the recovery observation can
be barely made since the recovery level is not large and 
we can't exclude that the afterglow can cause some average shift
of the signal at the first measurement.

\section{ Change of the crystal light output nonuniformity after
irradiation }

\begin{figure}[hbtp]
  \centering
  \includegraphics[width=0.7\textwidth]{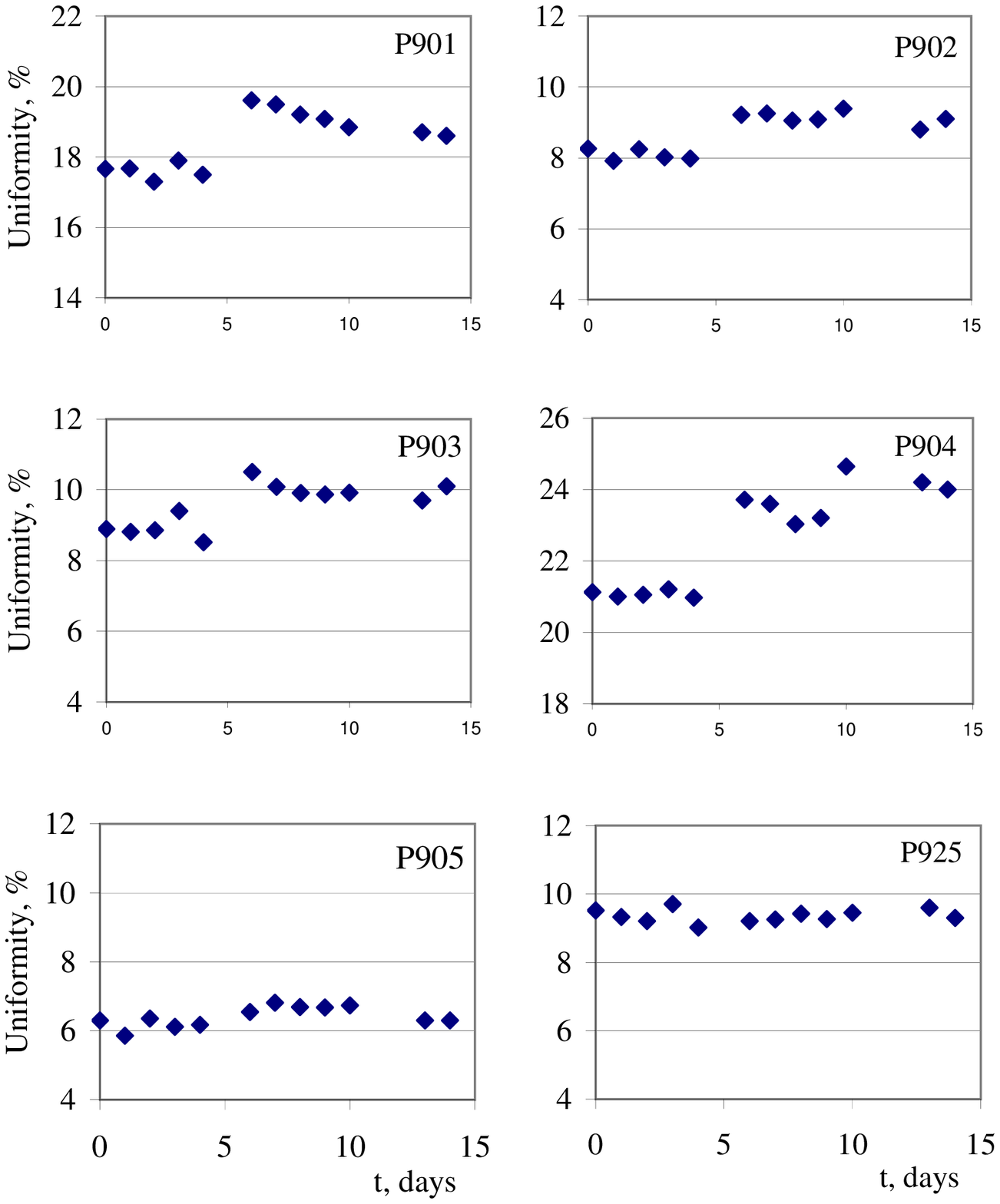}
  \caption{Time dependence of the nonuniformity of crystals.
  The step is connected to the irradiation with the 
  dose of 3200rad.
  Results with non irradiated crystal P925 are referred
  for comparison.}
\label{uniform}
\end{figure}

Some change of nonuniformity was observed after irradiation with a
dose of 3200~rad. 
For example, Fig.~\ref{uniform} shows the nonuniformity time 
dependence for several crystals of "P" type. Certain deterioration
of light output uniformity is seen for all irradiated crystals
after dose of 3200~rad. The nonuniformity of the sample P925
which was not irradiated referred for a comparison.
The value of the crystal nonuniformity was defined by the
formulae (\ref{nonuni}). 

\section{ Discussion } \label{dis}

Most of the papers devoted to a study of radiation hardness of 
scintillation crystals consider the optical transparency 
deterioration as a main reason of light output loss after
the irradiation (see, for example, \cite{zhu98,rad99}).
On the other hand, some authors presented mechanism of 
the damage of intrinsic scintillation efficiency of
the material due to irradiation \cite{rad-sc1,rad-sc2}.
The detail discussion of the physical models of the
radiation induced changes of the  properties
of the scintillation crystals as well as the wide
bibliography on this subject can be found in the book \cite{grin}.

In this work we did not make special studies of the 
mechanism of the light output degradation. It's worth to
remark, however, that during all of the time of this work 
the dose intensity were measured with
only one small CsI(Tl) crystal ($2\times 2\times 1cm^3$) which were
calibrated time to time. The total dose absorbed by this crystals
amounted to 50 krad. Nevertheless, we did not observe 
considerable decrease of light output of this crystal within few
percent of accuracy of measurements .
In contrast, the light output loss of the large crystals 
occur at the much lower absorbed dose that can be explained
assuming that the increasing light absorbtion provides the
main contribution to the light output degradation.
     
In general, the inference that the main radiation stimulated effect
is the loss of optical transparency of the crystal, looks to be
well-grounded, at least for moderate doses of radiation.
Following arguments provide evidence for this:
\begin{itemize}
  \item 
  the radiation induced absorbtion are really observed in the
  direct measurements of the optical transparency of crystals
  \cite{tit-rad,shpil} and its quantity does not contradict
  to observed light output decrease for the detectors of real
  sizes;
  \item 
  crystals of small sizes demonstrate much lower degradation
  of light output than the large crystals at the same dose.
\end{itemize}

It should be noted that besides the loss of the optical
transparency, the irradiation can produce both elastic and 
inelastic centers of photon scattering
in the material.
This effect can provide the deterioration of the light
collection efficiency as well. However, this effect was
not carefully studied for CsI crystals.

Let's discuss the results obtained in the sections \ref{me_piram}
and \ref{me_paral}.
Here, first of all, two features attract an attention:
a) large spread of the light output loss values after irradiation for
different samples of the same type and 
b) considerable average difference in $\Delta{L}/L$ between samples
having the shape of truncated pyramid and parallelepiped.

The conclusion coming from a) is the radiation hardness of the 
crystals depends strongly from uncontrolled variations in the
growing technology from one ingot to the another one. 
The high correlation between the radiation hardness of the
samples produced from the same ingot is the evidence of small
variations of the conditions during the growing process.
When the radiation hardness is a critical parameter of the detector,
the tests for at least one sample cut from each ingot should be
performed.

Then we would like to refer some considerations which can 
explain the effect b).
First of all, the light collection conditions for "B" and "P"
samples should be considered.
The sensitive area of PM tube photocathode, $S_{ph} \approx$~
16 cm$^2$, covers completely the output edge of "P" type crystals
$S^P_{out}$=4 cm$^2$, while this coverage is only partial for
crystals of type "B", $S^B_{out}$= 35$ \div$47 cm$^2$, 
$S_{ph}/S^B_{out}$ = 0.35$\div$0.44.
Meanwhile, the average light output for crystals of "B" type is 
only 15\% less than that for "P" crystals. 
One can suppose that the light collection in "B" crystals
involves substantially photons which had multiple reflection at
the crystal sides and, possibly, the scattering in the 
crystal volume.
Besides, the shape of the crystal plays substantial role for the
light collection. These effects should lead to the increase
of the photon mean path before it hits PM tube and, respectively,
to the higher sensitivity to the loss of the 
optical transparency.

\begin{figure}[hbtp]
  \centering
  \includegraphics[width=0.48\textwidth]{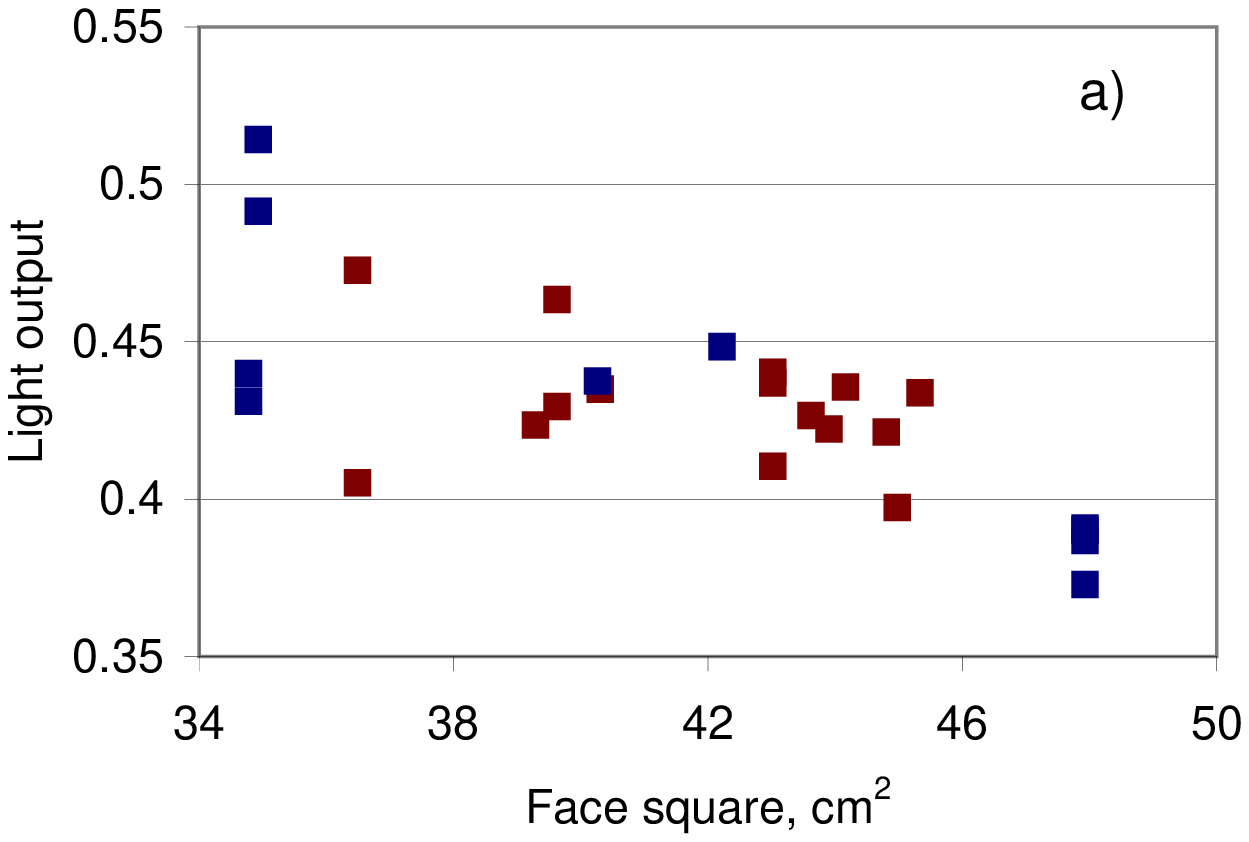}
  \hfill
  \includegraphics[width=0.48\textwidth]{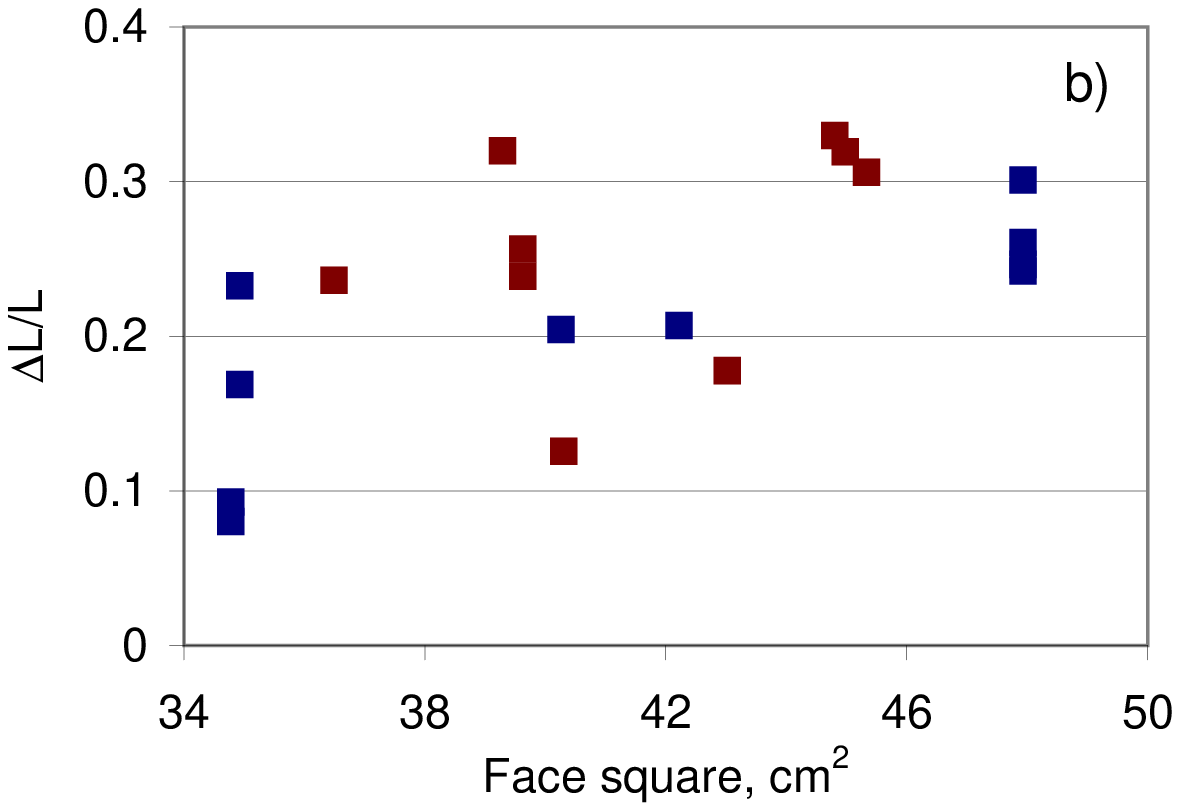}
  \caption{{\bf a)} -- initial light output of the crystals of
  "B" type in dependence of the output square;  {\bf b)} --
  the light output loss for samples of "B" type irradiated with
  a dose of 3200-3700 rad versus of the square of output face.}
\label{square}
\end{figure}

The initial values of the light output and the values of $\Delta{L}/L$
for 19 irradiated crystals of "B" type versus the square, $S$, of 
the output crystal face are presented  in Fig.~\ref{square}.
Some correlation between light output and $S$ as well as
between $\Delta{L}/L$ and $S$ can be seen in spite of the large
spread of data.                

\section{ Conclusion}

In conclusion we list the main results of this study:

\begin{itemize}
  \item 
  In general, all studied crystals have high enough radiation 
  hardness to be used in the detectors for existing B-factories;
  \item 
  Crystals produced within one technology using the raw material
  from the same manufacturer demonstrate, nevertheless,
  large spread of values of radiation stimulated light output
  loss; 
  \item 
  However, crystals cut from the same ingot have the same level of 
  radiation hardness;
  \item 
  Large difference in the values of radiation induced light collection
  efficiency loss for the samples of different geometry was observed.
  This is an evidence of the substantial dependence of the radiation
  hardness on the shape of the crystal and light collection
  conditions.   
  \item 
  There are some samples with high (for alcali halide scintillation
  crystals) radiation hardness   ($\Delta{L}/L$=8\% at 3200 rad of
  absorbed dose). This confirms a principle possibility of the
  mass production of radiation hard crystals after certain
  technology upgrade.  
  
\end{itemize}

{\bf Acknowledgements}

We are grateful to Prof.R.A.Salimov and Prof.N.K.Kuksanov
for the opportunity to use the ELV-6 accelerator. We would
like to thank Prof.A.E.Bondar for multiple fruitful and
helpful discussions.

\newpage

\renewcommand{\appendixname}{Appendix}

\appendix
\section{Absorbed dose measurement}

The detector of radiation dose in this work based on CsI(Tl) crystal.
This detector of $20\times 20\times 10$~mm$^3$ size 
was covered with two layers of the
200 $\mu m$ porous teflon tape. The light signal is read out 
with a vacuum photodiode (VPD) with massive photocathode. The bias
voltage of 40 V is applied as it is shown in Fig.~\ref{izmer}.
The current trough the photodiode is measured.

The choice of the VPD with the solid photocathode is caused by 
the reason that the measured current can be rather high, up to
10--20 $\mu A$. The vacuum photodetectors with semitransparent 
photocathodes loose the efficiency when the photo current 
exceeds $\sim10 nA$ due to the disturbance of the electric
potential over the cathode surface.

The another option for the scintillation light read out is
the usage of the silicon photodiode. But in this case
one must take into account the current induced by the
interaction of gamma radiation with silicon in addition 
to the photo current that can decrease the accuracy of
the measurement.

The dose rate is determined by the expression:
\begin{equation}\label{dDdt}
  \frac{dD}{dt} = I_D/I_R ,
\end{equation}
where $I_D$ --- measured photo current, and $I_R$ is the calibration 
constant, that is equal to the photo current value at the dose rate of 
1 rad/sec.
\begin{equation}\label{IR}
  I_R = C\cdot m\cdot e \cdot N_e^M \cdot K_{\tau} ,
\end{equation}
where C=6,24$\times$10$^8$ MeV/g; $m$ is crystal mass in g; 
e is an electron charge, $ N_e^M$ -- normalization constant, 
the number of electrons passed trough photodiode per 1 MeV of 
the energy absorbed by CsI;
$ K_{\tau} $ is a correction factor that will be explained 
later.

Thus, the key task for the described method of the dose measurement
is the correct calibration, i.e. a determination of the $ N_e^M$ 
value and, with its help, the value of $I_R$.

\begin{figure}[hbtp]
\begin{center}
\includegraphics[width=100mm] {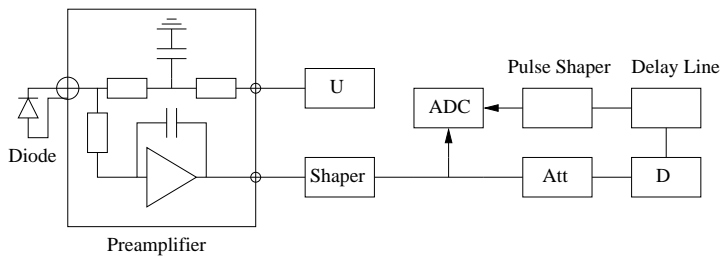}
\caption{The layout of the dose detector calibration.
Diode~---~silicon or vacuum photodiode,
Preamplifier~---~charge sensitive preamplifier,
U~---~voltage source, Att~---~attenuator, D~---~discriminator.}
\label{cal_scheam}
\end{center}
\end{figure}

The layout of the DD calibration is presented in Fig.~\ref{cal_scheam}

A photodiode is connected to the input of the charge sensitive
preamplifier (CSPA). The output signal comes to the shaping
amplifier with the shaping time of 2~$\mu s$. The shaped
pulse is digitized by the ADC.

At the first stage of the calibration the ADC scale constant
was determined, i.e. the number of electrons at the CSPA input
pulse that corresponds to one ADC bin. To do that the CSPA input
was connected to the silicon PIN photodiode irradiated with 
$\gamma$-quanta of 60 keV energy from  $^{241}Am$ source.
The calibration constant was determined as:
\[ \kappa = \frac{E_{\gamma}}{E_i}\cdot \frac{1}{A_{peak}} , \]
where $A_{peak}$ is a peak position in the pulse height spectrum, 
$E_i$ --- the average ionization energy for silicon,  
$ E_i= $~3.62 eV. 

\begin{figure}[hbtp]
\begin{center}
\includegraphics[width=100mm] {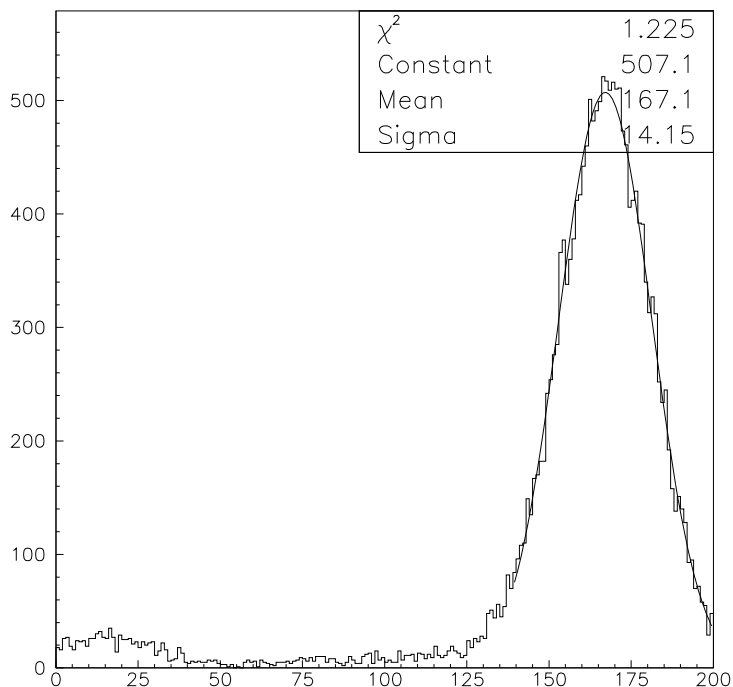}
\caption{$^{238}Pu$ pulse height spectrum.}
\label{sppu}
\end{center}
\end{figure}

To measure the signal amplitude corresponding to the 1 MeV 
energy absorbed in the CsI(Tl) crystal the $\alpha$-source
$^{238}Pu$ ($ E_{\alpha}=5.1~MeV $) was used.
The direct measurement of this value with 
$\gamma$-source $^{137}Cs$ (662 keV) or $^{22}Na$ (511~keV and
1275~keV) was not possible since the noise level was found 
to be equivalent of 300 keV. To avoid the $\alpha$-particle
absorption in the teflon coverage of the crystal, the
small hole of about 1 mm$^2$ was made in that. The measured
pulse height spectrum is shown in Fig.~\ref{sppu}.

It's known that the scintillation light output for $\alpha$-particle 
is lower than
that for photons or electrons due to much higher ionization density. 
The ratio of the signals from $ ^{238}Pu $ and total absorption 
peak of $ ^{137}Cs $ was measured separately with the light
read out with PM tube. For the crystal used in DD 
it was found that peak of $ ^{238}Pu $ corresponds to  
$ 3.71\pm 0.09$~MeV of photon energy.

The resulting number of photo electrons per 1 MeV of the 
energy deposited in the crystal was measured to be 
$N_e^M=(1180 \pm 30)$~p.e./MeV.
It should be noted that for ADC scale calibration the short
pulse (about 10 ns) from PIN photodiode was used while 
the scintillation flash of CsI(Tl) has much longer decay time.

To take this effect into account, the correction factor, 
$ K_{\tau} $, was implemented to the formulae (\ref{IR}).
Its value was calculated using the light pulse shape 
measured in \cite{gar} where the experimental shape of this
pulse was well approximated by the sum of three exponents
with the decay constants of $\tau_1= 0.7 \mu s$, 
$\tau_2=2.5 \mu s$ and $\tau_3=17 \mu s$.
As a Laplas representation of the shaper (ORTEC 370)
response function,
the following expression was used:
\[ K_L(p) = \frac{p}{\tau_s^2(p+1/\tau_s)^3} ,\]
where  $\tau_s$ is the shaping time.
This formula implied the single differentiation and 
double integration provides good approximation of the
output pulse shape that corresponds to the input step
pulse. The measured value of the correction factor 
was $ K_{\tau} =1.4$.

As a result of the described calibration procedure the
value of $I_R = 0.3 \mu A\cdot$s/rad was obtained.
This procedure was repeated several times with the
interval of several months giving the close results of $I_R$.

We estimate the relative error of the absorption dose
measurement with the described method by about 10\%.
The main contributions to the inaccuracy are connected with
the inexact description of the slow components of 
scintillator decay time as well as with precision of 
the measurement of $\alpha$/$\gamma$ ratio.

 


\newpage

\end{document}